\documentclass[aps,prl,reprint,twocolumn,groupedaddress,showpacs]{revtex4}

\usepackage{hyperref}

\usepackage{amsmath}
\usepackage{amsfonts}
\usepackage{amssymb}
\usepackage{graphicx}
\usepackage{empheq}

\hypersetup{colorlinks=true, urlcolor=blue, linkcolor=blue, citecolor=blue,plainpages=false} 

\newcommand{\tr}{\mbox{tr}}
\newcommand{\diff}[2]{\frac{d{#1}}{d{#2}}}

\begin{document}
\title{Optimal Frames for Polarisation State Reconstruction}

\author{Matthew R. Foreman$^{1}$}
\email[]{matthew.foreman@mpl.mpg.de}
\author{Alberto Favaro$^2$}
\author{Andrea Aiello$^{1}$}
\affiliation{$^1$Max Planck Institute for the Science of Light,
G\"unther-Scharowsky-Stra{\ss}e 1, 91058 Erlangen, Germany\\
$^2$Blackett Laboratory, Department of Physics, Imperial College
London, Prince Consort Road, London, SW7 2AZ, UK}

\begin{abstract}
Complete determination of the polarisation state of light requires
at least four distinct projective measurements of the associated
Stokes vector. Stability of state reconstruction, however, hinges on the condition number $\kappa$ of
the corresponding instrument matrix. Optimisation of redundant
measurement frames with an arbitrary number of analysis
states, $m$, is considered in this Letter in the
sense of minimisation of $\kappa$. The minimum achievable $\kappa$ is
analytically found and shown to be independent of
$m$, except for $m=5$ where this minimum is 
unachievable. Distribution of the optimal analysis states over the
Poincar\'e sphere is found to be
described by spherical 2-designs, including the 
Platonic solids as special cases. Higher order polarisation
properties also play a key role in nonlinear, stochastic and
quantum processes. Optimal measurement schemes for nonlinear measurands of degree $t$ are hence also considered and found to correspond to
spherical $2t$-designs, thereby constituting a generalisation of the concept of mutually unbiased bases.
\end{abstract}



\maketitle

Measurement of the polarisation of light is a common problem in many fields of
physics including quantum information, astronomy, quantitative biology
and single molecule orientational imaging
\cite{Salvail2013,Hadamcik2010,Sofikitis2014,Foreman2008c}.  Typically, determination of the polarisation state of light, as
parameterised by a  $4\times 1$ Stokes vector, $\mathbf{S}$, follows by
making projective measurements onto a set of known analysis states, with complete state
reconstruction requiring a minimum of four distinct
measurements \cite{Damask2005}. Arguably the simplest so-called complete polarimeter is that
which comprises of linear polarisers, oriented at $0^\circ$, $45^\circ$
and $90^\circ$ to some reference axis, and a circular polariser. By virtue of the linear nature of the
measurement process, the intensities transmitted through each
polarisation state analyser (PSA), denoted by $\mathbf{D}$, can be related to the
incident Stokes vector via $\mathbf{D} = \mathbb{A} \mathbf{S}$, where
$\mathbb{A}$ is known as the instrument matrix (in this example
$\mathbf{D}$ is $4\times 1$ and $\mathbb{A}$ is $4\times 4$ in dimension). 
Although intuitively simple to understand and easy to implement, this polarimeter performs
sub-optimally. Much effort has been invested over the years to
optimise the geometry of polarimeters using metrics such as the total 
variance on the inferred Stokes vector \cite{Sabatke2000,Goudail2009},
information content \cite{Foreman2008b,Foreman2010,Tyo1998a,Alenin2015}, the
determinant of the instrument matrix
\cite{Azzam2003,Ambirajan1995a,Ambirajan1995b} and signal to noise ratio
\cite{Tyo2002a}. 
Perhaps most popular,
however, is the condition number, $\kappa$, of the instrument
matrix \cite{Ambirajan1995a,Ambirajan1995b,Azzam1988,Tyo2002a,Savenkov2002,Smith2002a,Twietmeyer2008,Peinado2010}
which describes the stability of the polarisation inference problem
regardless of reconstruction algorithm and bounds the
extent to which relative measurement errors are amplified during state reconstruction.  Smaller
condition numbers imply more robust measurements. 
Use of a measurement set of greater than four analysis states, however, is known
to mitigate the effects of noise
\cite{Aiello2006,Foreman2008b,Peinado2010}. Nevertheless, only limited consideration has been given to optimisation of these systems. Drawing from results in discrete
computational geometry, linear algebra and
state tomography this work hence considers
the optimisation of such measurement schemes in the sense of
minimisation of $\kappa$ and
discusses the associated geometric interpretation. It is established
analytically that the minimum condition number is $\sqrt{20}$
independent of the number of analysis states. Formal equivalence between minimisation of $\kappa$, maximisation of the determinant of the associated Gram
matrix and minimisation of the equally weighted variance is
established. Optimality constraints
are further derived and used to construct some illustrative optimal
measurement sets. More specifically, it is found that the distribution
of optimal analysis states over the Poincar\'e sphere is intimately
related to spherical 2-designs. Accordingly, in contrast to previous reports, optimal
measurements are shown not to necessarily correspond to inscribed polyhedra of
maximal volume.

Going beyond linear reconstruction of the Stokes vector,
measurements of more complex functions of the Stokes vector,
$D(\mathbf{S})$, are made in a number of applications. Nonlinear light
scattering and material characterisation, for example, gives rise to
intensities which depend on products of Stokes
parameters \cite{Shi1994,Samim2015}. Furthermore, random media
and rough surfaces can be studied through the changes in the
statistical properties of the polarisation of light induced upon transmission or
reflection \cite{Barakat1987a,Aiello2005,Gil2007a}. Full
characterisation, however, requires determination of the underlying
probability distribution function, or equivalently all higher order statistical moments of the Stokes parameters. Optimal
analysis states for such higher order problems are also considered in this
Letter and their relationship with spherical $t$-designs established. Importantly, although the language of classical polarimetry
will be used throughout, it should be noted that the results given are equally applicable in the quantum regime for states of
a given number of photons. For example, knowledge of higher order
moments of the Stokes operators can give insight into hidden ``quantum'' polarisation
in a classically unpolarised state \cite{DeLaHoz2013,Soderholm2012}.

\begin{figure*}[!t]
\begin{center}
	\includegraphics[width=0.95\textwidth]{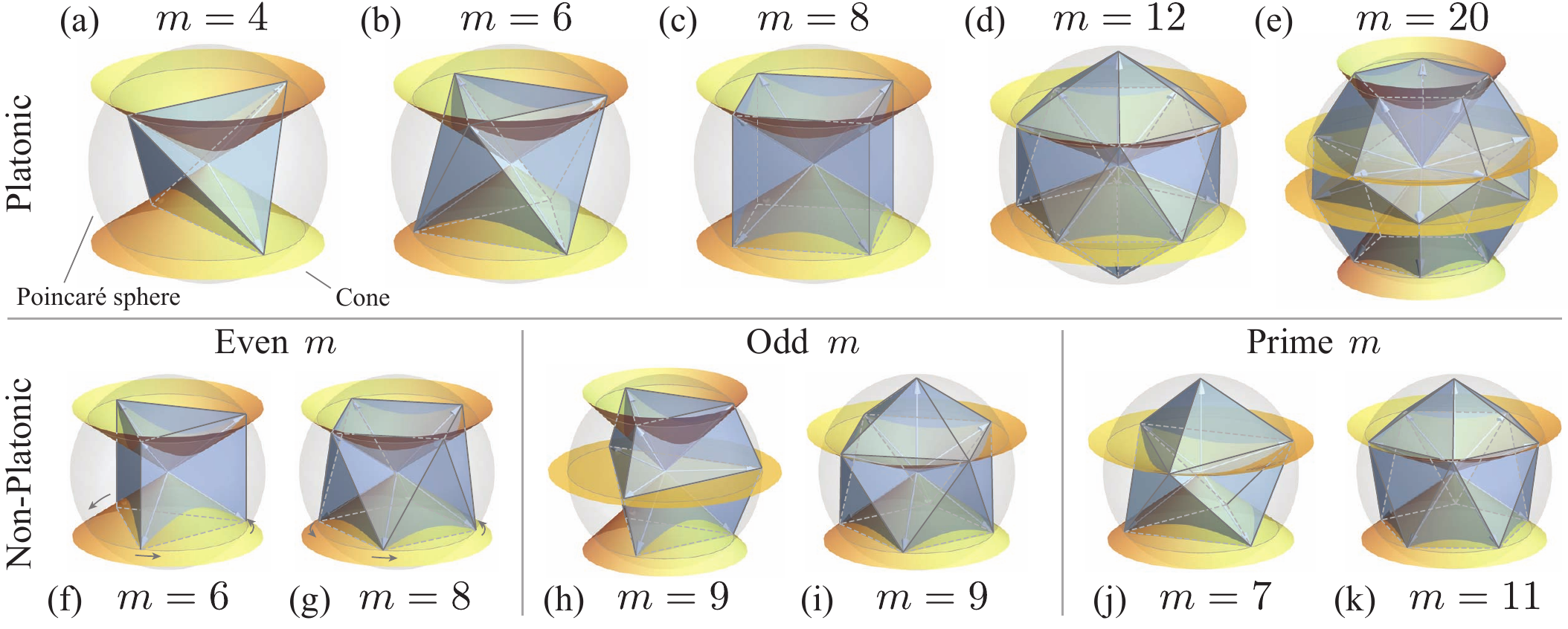}
	\caption{(a)-(e) Optimal measurement frames for
          $m=4,6,8,12$ and 20 defining the Platonic polyhedra inscribed in
          the Poincar\'e sphere. Corresponding analysis states lie on
          the family of cones shown. (f)-(k) Example optimal measurement
          frames and associated non-Platonic polyhedra for $m= 6,8,9,7$
          and 11.  \label{fig:polyhedra}}
\end{center}
\end{figure*} 

The action of a PSA on incident light of
arbitrary polarisation, can be considered as a projective
measurement, whereby the output light has polarisation matching the
nominal analysis state of the PSA (denoted $\mathbf{A} =
(1,A_1,A_2,A_3)^T/2$ with $\sum_{k=1}^3 A_{k}^2 = 1$) and
intensity of $\mathbf{A}^T \cdot \mathbf{S}$, where $\mathbf{A}$ has been normalised to ensure the PSA
is passive \cite{Foreman2010}. Each row of the instrument matrix
associated with $m \geq 4$ different measurements is given by
$\mathbf{A}_j$ ($j=1,2,\ldots,m$) and $\mathbb{A}$ thus has dimension
$m\times 4$. The corresponding vector of measured intensities,
$\mathbf{D}$, is hence $m\times 1$. Geometrically, each $\mathbf{A}_j$ can be considered as
defining a point on the surface of the Poincar\'e sphere through the
reduced vector $\mathbf{a}_j =
(A_{j1},A_{j2},A_{j3})^T $, such that $\mathbb{A}$ defines the vertices of a
polyhedron inscribed within the unit sphere. Moreover, the set of vectors
$\{\mathbf{a}_j\}$ constitutes a measurement frame \cite{Tyo2002a}.

The
condition number of the instrument matrix  is explicitly defined as $\kappa = \|\mathbb{A}\| \| \mathbb{A}^+ \|$, where $\mathbb{A}^+$ denotes the generalised inverse of $\mathbb{A}$ 
and $\| \cdot \|$ denotes the matrix norm (taken as the
Hilbert-Schmidt norm throughout this work). Noting that the normalisation
imposed on the rows of $\mathbb{A}$ requires
$\mathbf{A}_j^T\cdot\mathbf{A}_j = 1/2$, it follows that 
\begin{align}
\| \mathbb{A} \|^2 = \tr[\mathbb{A}^T \mathbb{A} ] =m/2 \label{eq:normA}.
\end{align}
For any given experimental setup $\|\mathbb{A}\|$ is thus
constant. Accordingly, minimisation of the condition number is
equivalent to minimisation of $\|\mathbb{A}^+\|^2$, an alternative figure of merit known as the equally weighted
variance (EWV) \cite{Sabatke2000,Peinado2010}. The EWV quantifies the
noise amplification in the reconstructed Stokes vectors
assuming a least norm reconstruction and equal magnitude errors on
each measurement \cite{Sabatke2000}. Similarly to above $\| \mathbb{A}^+ \|^2  =\tr[(\mathbb{A}^+)^T \mathbb{A}^+]
=\tr[\mathbb{B}^{-1} ]$, 
where  $\mathbb{B} = \mathbb{A}^T \mathbb{A}$ is a $4\times 4$ 
matrix and $\mathbb{B}^{-1}$ denotes its inverse. 
The inverse of $\mathbb{B}$ exists if $\mathbf{a}_j$ are
not all co-planar and moreover can be written in the
form $\mathbb{B}^{-1} =  \mbox{adj}[\mathbb{B}]  / |\mathbb{B}|$ 
where $|\cdot|$ and $\mbox{adj}[\cdot]$ denote the determinant and 
adjunct of a matrix respectively. The condition number ($\kappa > 0$)
can thus be expressed as
\begin{align}
\!\kappa^2 = \tr[\mathbb{B}]\tr[\mathbb{B}^{-1} ]=
\frac{1}{2}\frac{m}{| \mathbb{B} |}
\tr\left[\diff{|\mathbb{B}|}{\mathbb{B}}\right]
=  \frac{m}{2} \sum_{i = 1}^{4} \diff{\,\ln |\mathbb{B}|} {B_{ii}} \label{eq:kappa2}
\end{align}
where Jacobi's formula has been used
\cite{Prasolov1994} 
and $B_{ij}$ is the $(i,j)$th element of $\mathbb{B}$. Upon differentiation of Eq.~\eqref{eq:kappa2}, application of the product
rule and back-substitution it can be shown that $2\,d \ln \kappa = -d  \ln |\mathbb{B}|$.
Minimisation of the
condition number of the instrument matrix is hence also equivalent
to maximisation of the determinant of $\mathbb{B}$. Geometrically, it
is interesting to note that $|\mathbb{B}|$ represents
the volume squared of a 4-parallelotope in $\mathcal{R}^m$.  For the special
case of $m=4$, maximisation of $|\mathbb{B}|$  is
equivalent to maximising the volume of
the tetrahedron whose vertices in $\mathcal{R}^3$ are defined by
$\mathbf{a}_j$  (see Fig.~\ref{fig:polyhedra}(a)) as has
been previously reported \cite{Azzam1988,Savenkov2002}.

Hadamard's inequality \cite{Prasolov1994} states that the determinant
of $\mathbb{B}$ is upper bounded by the product of its
diagonal elements. The maximum determinant is thus obtained when
$\mathbb{B}$ is diagonal whereby the diagonal elements also correspond to the eigenvalues $\beta_l$
($l=1,\ldots,4$). Explicitly $\mathbb{B}$ can be expressed in the form
$\mathbb{B} = \sum_{j=1}^m \mathbf{A}_j \mathbf{A}_j^T$, whereby it
follows by inspection that $B_{11} = \beta_1 = m/4$. Furthermore
maximisation of $|\mathbb{B}|$ is subject to the constraint
$\tr[\mathbb{B}] = m/2$ (c.f. Eq.~\eqref{eq:normA}). 
Use of the method of Lagrange multipliers then yields
$\beta_1 /3 =\beta_2 = \beta_3 = \beta_4
= m /12$, i.e. the condition number of $\mathbb{A}$ is
minimised when
\begin{align}
\mathbb{B} = \frac{m}{12}\left(\begin{array}{cccc} 3 & 0 & 0 &0 \\
0 & 1 & 0 &0 \\
0 & 0 & 1 &0 \\
0 & 0 & 0 &1   \end{array}\right). \label{eq:Bform2}
\end{align}
Substitution of Eq.~\eqref{eq:Bform2} into Eq.~\eqref{eq:kappa2} gives
$\kappa = \sqrt{20}\approx 4.47214$. The condition number of an
optimised polarimeter is hence independent of the
number of analysis states $m$. It
also follows that $|\mathbb{B}| =  m^4 / 6912$ and the EWV $=
40/m$. The fall off in the EWV with increasing number
of measurements reflects the noise reduction arising from greater
measurement redundancy.

Generalisation of the above
to the problem of reconstruction of an $N \geq 3$ dimensional vector
$\mathbf{S}_N = (S_0,S_1,\ldots,S_{N-1})^T$, for which $S_0^2 \geq \sum_{j
  = 1}^{N-1} S_j^2$, using $m \geq N$ projective measurements onto
analysis states of the form $\mathbf{A}_{N} =
(1,A_1,\ldots,A_{N-1})^T/2$ can also be easily performed. Consideration of the $N=3$
(i.e. reduced dimensionality) case is applicable to 
linear polarimetry for example, whereas higher
dimensional generalisations
are relevant to polarimetry of three dimensional fields, whereby Stokes vectors become $9\times 1$ in size
\cite{Roman1959,Gil2007a}. Following the steps given above in the $N$
dimensional case it is found that $|\mathbb{B}|$ is maximised when $\mathbb{B}$
is diagonal with non-zero elements of $\beta_1 = m / 4$ and
$\beta_l = \beta_1 / ( N -1)$ for $l \neq 1$. In turn it follows that
$|\mathbb{B}| = (m/4)^N (N-1)^{1-N}$,  $\kappa^2 = 2 N^2 - 4 N + 4$
and the $\mbox{EWV} = 2 \kappa^2/m$.

Whilst the above treatment has considered the minimum achievable
condition number and EWV, the optimal measurement basis has not yet
been determined. To this end, Eq.~\eqref{eq:Bform2} must be invoked
which upon generalisation implies the set of polynomial constraints
\begin{align}
\sum_{j=1}^m \mathbf{a}_j = \mathbf{0} &&\mbox{and}&& \sum_{j=1}^m
\mathbf{a}_j\mathbf{a}_j^T = \frac{m}{N-1} \mathbb{I}_{N-1}, \label{eq:sol_set}
\end{align}
where $\mathbb{I}_N$ is the $N\times N$ identity matrix. A
measurement frame is optimal iff Eq.~\eqref{eq:sol_set}
is satisfied. When $N=3$ it can be shown \cite{Haase1996} that the optimal
measurement frame corresponds to $\mathbf{a}_j$ defining a regular polygon inscribed in
a unit circle. Incidentally, the regular inscribed polygons have the maximum
area of all inscribed polygons. 
For the $N=4$ case (which is exclusively considered henceforth), it is
found that Eq.~\eqref{eq:sol_set} is satisfied if the set of vectors $\{
\mathbf{a}_j \}$, or equivalently the vertices of the underlying polyhedron, constitute a spherical 2-design
in $\mathcal{R}^3$ and are thereby closely related to mutually
unbiased bases \cite{Klappenecker2005a}. Proof of this result follows from the
definition of spherical
$t$-designs
as a collection of $m$ points on the surface of the unit sphere
in $\mathcal{R}^3$ for which the (normalised) integral of any
polynomial, $g(\mathbf{S})$, of degree
$t$ or less is equal to the average taken over the $m$
points \cite{Delsarte1977,Mimura1990}. Use of the
polynomial functions $g = S_j$ and $S_j S_k$ ($j$ and $k =1,2,3$)
in this definition yields Eq.~\eqref{eq:sol_set}. It is also
worthwhile noting that a spherical
$t$-design is also a $t-1$ design \cite{Delsarte1977}. 

Numerical codes can be used for determination of spherical
$t$-designs in general \cite{Hardin1996}, however, in view of the symmetry of the $N=3$ solution, an analogous symmetry in the
$N=4$ case is expected and can be used to guide the construction of some
simple $2$-designs. Specifically, letting
$m=rs$, and adopting the initial ansatz $\mathbf{a}_j = (\sin\theta_q
\cos [\phi_p + \Phi_q] , \sin\theta_q\sin [\phi_p+\Phi_q] , \cos\theta_q )^T$
for $j = 1,\ldots ,m$, $p = 1,\ldots, r$, $q = 1,
\ldots, s$, where $\phi_p = 2\pi \, p / r$, Eq.~\eqref{eq:sol_set} reduces to
\begin{align}
\sum_{q=1}^{s} \cos\theta_q = 0 && \mbox{and}&
&\sum_{q=1}^{s}  \cos^2\theta_q= \frac{s}{3} .\label{eq:f_sol}
\end{align}
In general, the solution to Eq.~\eqref{eq:f_sol}, and hence the choice
of optimal measurement frame, is not unique (even allowing for the intrinsic rotational freedom). Nevertheless, a number of 
solutions can be found as is now illustrated. 

\textbf{Even $m$:} For even $m$ a simple optimal frame 
follows by taking $s=2$ whereby Eq.~\eqref{eq:f_sol} is trivially satisfied when $\cos\theta_1 =
-\cos\theta_2 = 1/\sqrt{3}$. Without loss of generality $\Phi_1$
can also be set to zero. If $\Phi_2 = 0$ it  follows that, depending on the relative sign of $\sin\theta_1$ and
$\sin\theta_2$, optimal frames possessing either an inversion or
mirror symmetry can be generated. In both cases, however, the vectors $\mathbf{a}_j$ lie
on two cones with apex at the origin and apex
angle $2\theta_q$ (see e.g.,
Figs.~\ref{fig:polyhedra}(b), (c), (f) and (g) for $m=6$ and 8). Geometrically, taking $\Phi_2
\neq 0$ corresponds to a rotation of the measurement states on the
lower cone relative to those on the upper cone as depicted in
Figs.~\ref{fig:polyhedra}(f) and (g). Solutions
describing the regular tetrahedron $(m,r,s) = (4,2,2)$, octahedron
$(6,3,2)$ and cube $(8,4,2)$ can be generated in this manner (see Figs.~\ref{fig:polyhedra}(a)-(c)). 

Going further, a measurement frame containing $m$ analysis states
can be partitioned into $M$ subsets of size $m_i = 2\mu_i$, where $m
= 2 \sum_{i=1}^M \mu_i \equiv 2 \mu$. Optimal bases can then be constructed
($\Phi_q = 0$ is assumed henceforth for simplicity) by
constraining the vectors of each subset to lie on cones with coinciding axes,
 but with differing apex angles. Explicitly it can then be shown that
 the apex angles $\theta_i$ ($i=1,\ldots,M$) of each cone are given by $
\sin^2\theta_i = {2\mu  \lambda_i}/{(3\mu_i)}$,
where $\sum_{i=1}^M \lambda_i = 1$ \cite{Haase1996}. Note that in the
limiting case of $\theta_i \rightarrow 0$ (whereby necessarily $\mu_i=1$), the corresponding cone
collapses to its axis (e.g. Fig.~\ref{fig:polyhedra}(d)). 
Both the regular icosahedron and dodecahedron can hence be constructed
(Figs.~\ref{fig:polyhedra}(d) and (e)). Possible optimal measurement
frames for the $m = 4,6,8,12$ and
 $20$ cases are thus defined by the vertices of the
 Platonic solids inscribed in the Poincar\'e sphere, in agreement with the numerical results of \cite{Peinado2010}. The inscribed polyhedra generated for $m=4, 6$
 and 12 correspond to the polyhedra of maximal volume. In contrast,
 however, noting that inscribed polyhedra of maximal
 volume are Euclidean simplexes \cite{Berman1970},
i.e. each face is triangular, the cube and  dodecahedron do not have maximum volume in contrast to the claim of \cite{Peinado2010}. Non-uniqueness of the solution to
 Eq.~\eqref{eq:sol_set} implies the
 associated polyhedra for arbitrary $m$ are also not of maximal volume in general.

\textbf{Odd $m$:} When $m$ is a factorable odd integer the ansatz used
thus far is also capable of generating optimal measurement 
frames. In Fig.~\ref{fig:polyhedra}(h), for
example, a possible optimal frame is shown for $(m,r,s)=(9,3,3)$, with
$\cos\theta_2 = 0$ and $\cos\theta_{1,3} = \pm 1/\sqrt{2}$. For prime
$m$ alternative solutions must, however, be sought. It is well known
that no spherical $2$-design, and hence no optimal frame, exists for
$m=5$ \cite{Mimura1990}, as can be verified by calculation of the Groebner
basis \cite{Cox2007} of the set of polynomial equations given by
Eqs.~\eqref{eq:sol_set} and the constraints $\mathbf{a}_j^T \cdot \mathbf{a}_j = 1$. Solutions for larger
prime $m$ can nevertheless still be found. One such solution set (also valid for
factorable odd $m$)
takes the form $\mathbf{a}_j = (\sin\theta_j \cos \phi_j ,
\sin\theta_j\sin \phi_j,\cos\theta_j)$,
where $\phi_j = 2\pi j / (m-1)$, $\cos\theta_j = c_1$ for $j =
1,3,\ldots, m-2$, $\cos\theta_j = c_2$ for $j =
2,4,\ldots, m-1$, $\cos\theta_m = 1$ and 
$c_i = [{3 \pm (-1)^i \sqrt{3 m (m-4)}}]/{[3(1-m)]} $ as follows from Eq.~\eqref{eq:sol_set}.
This solution is depicted in Figs.~\ref{fig:polyhedra}(i)-(k) for $m= 7, 9$ and 11.

The underlying mathematical framework found above hints at a potential extension of the
optimisation procedure to higher order measurement problems. Consider,
for example,
the measurement of a nonlinear function $D(\mathbf{S})$, such as the
generated intensity in an optical nonlinear conversion
problem. Letting $P_k(\mathbf{S})$ denote a complete basis
of polynomial functions, ordered by increasing polynomial degree and
indexed by $k = 0,1,\ldots, \infty$, the measurand can be
decomposed according to 
\begin{align}
D(\mathbf{S}) = \sum_{k=0}^\infty F_k P_k(\mathbf{S}) \label{eq:Dexpansion}
\end{align}
where $F_k$ are the associated expansion coefficients which are now
the unknowns of interest. $P_k$ are
assumed to be orthonormal over the Poincar\'e sphere. Practically, a finite number of measurements,
$m$, are made in directions $\{\mathbf{A}_j\}$ so as to sample
$D(\mathbf{S})$. Furthermore, the sum in Eq.~\eqref{eq:Dexpansion} must be truncated at a finite order
$K$, such that Eq.~\eqref{eq:Dexpansion} reduces to $\mathbf{D} = \mathbb{P} \,\mathbf{F}$, where $\mathbf{F}
= (F_0,F_1,\ldots,F_K)^T$ and $[\mathbb{P}]_{jk} =
P_k(\mathbf{A}_j)$. In general, at least $K+1$ measurements are required for
complete determination of $\mathbf{F}$. Optimality in this case can again be quantified
using the condition number of $\mathbb{P}$. Assuming the
polynomials $P_k$ satisfy $\sum_{k=0}^K
P_k^*(\mathbf{S})P_k(\mathbf{S}) = C$, where $C$ is a constant, a trace constraint
analogous to Eq.~\eqref{eq:normA} follows. Noting $P_0(\mathbf{S})$ is also
a constant, similar derivations to above yield 
\begin{align}
\kappa^2 = \frac{C}{P_0} + \frac{C K^2}{C - P_0^2},
\end{align}
where the corresponding optimal Gram matrix $\mathbb{P}^\dagger \mathbb{P}$ must be
diagonal. The resulting constraints (c.f. Eq.~\eqref{eq:sol_set}) are
satisfied if the analysis states constitute a spherical $2t$-design,
where $t$ is the degree of the polynomial $P_K$. Proof follows in a
similar fashion to above. As a concrete
example, if the polynomial functions $P_k$ are taken as the spherical
harmonics $Y_{lm}$ (whereby the $k$ index in Eq.~\eqref{eq:Dexpansion} denotes a suitable
lexicographic ordering of the indices $(l,m)$) up to maximum degree $l
= t$, it follows that $K =
(t+1)^2$ and $P_0 = 1/\sqrt{4\pi}$. Using the addition theorem $\sum_{m=-l}^l
Y_{lm}^*Y_{lm} = (2l+1)/(4\pi)$ the minimum
achievable condition number is found to be $\kappa =
(t+1)^2$. Importantly, it should be noted that spherical $2t$ designs
do not necessarily exist for arbitrary $m$ \cite{Hardin1996}. In
general it is therefore found that estimation of higher order
properties not only becomes more ill-conditioned, but also requires a
larger number of measurements to be made. Use of optimal measurement
sets is therefore critical to reconstruction quality in this case. In this sense the Platonic
solids perform well as measurement frames since they constitute higher
order $t$-designs, e.g. a regular dodecahedron is a spherical
$5$-design, whilst possessing a relatively small number of analysis states. 

In summary, optimal measurement frames for the
reconstruction of the Stokes vector $\mathbf{S}$ of polarised light have been
analytically and geometrically investigated. This analysis
can also be applied to the input polarisation states in Mueller matrix polarimetry \cite{Layden2012}. Equivalence of optimisation based on the EWV, the
condition number $\kappa$ of the associated instrument matrix and the
determinant of its Gram matrix was established. Constraints on the
optimal analysis states were derived and found to be satisfied by
states defining spherical  2-designs. It
 followed that minimisation of $\kappa$ does not necessarily correspond to maximisation of
 the volume of the corresponding inscribed polyhedron. Finally,
 results were extended to consider optimal frames when the measurand
 is a polynomial function of $\mathbf{S}$ of degree $t$. In this case optimal frames correspond to sets of analysis
 states constituting spherical $2t$ designs. This work provides the
 means for optimal polarisation state tomography and hence paves the
 way for practical study of nonlinear or stochastic properties of polarisation, which are of interest in both
 biological and physical contexts. Within the former, for
 example, nonlinear polarisation studies can provide insight into
cellular and molecular structure \cite{Mazumder2012,Samim2015}, whilst
the latter can enable 
study of fundamental quantum polarisation  properties
\cite{Soderholm2012,Klimov2010,Holleczek2011}. Finally, these results
also present interesting opportunities for establishing optimal 
schemes for quantum tomography of two level systems due to the underlying
geometric parallels with the Bloch sphere.

The authors acknowledge Markus Grassl for suggesting the
connection between the linear problem and spherical $2$-designs and for 
Luis L. S\'{a}nchez-Soto and Gerd Leuchs for insightful discussions. AF acknowledges financial support from the Gordon and Betty Moore Foundation.

\end{document}